# Size dependence of the pressure-induced phase transition in nanocrystals

Zhouwen Chen, Chang Q Sun,[a] and G. Ouyang
School of Electrical and Electronic Engineering, Nanyang Technological University, Singapore 639798
Yichun Zhou [b]
Key laboratory of low-Dimensional Materials and Application Technologies (Xiangtan University), Ministry of Education, Hunan 411105, China

Extending the recently-developed bond-order-length-strength (BOLS) correlation mechanism [Sun CQ, *Prog Solid State Chem* **2007,** 35, 1-159] to the pressure domain has led to atomistic insight into the phase stability of nanostructures under the varied stimuli of pressure and solid size. It turns out that the competition between the pressure-induced overheating ($T_C$ elevation) and the size-induced undercooling ($T_C$ depression) dominates the measured size trends of the pressure-induced phase transition. Reproduction of the measured size and pressure dependence of the phase stability for CdSe, $Fe_2O_3$, and $SnO_2$ nanocrystals evidences the validity of the solution derived from the perspective of atomic cohesive energy and its response to the external stimulus.

I       **Introduction**

Phase stability of nanostructures has been one of the central issues in nanoscience and technology, which has attracted considerable interest in recent years.[1,2,3,4,5,6] For a given specimen of a fixed size, phase transition takes place when the temperature is raised to a certain degree. The value of the critical temperature ($T_C$) varies with the actual process of phase transition. The $T_C$ values are different for the magnetic-paramagnetic, ferroelectric-non-ferroelectric, solid-solid, solid-liquid, and liquid-vapor phase transitions of the same specimen. Generally, the $T_C$ of a nanosolid drops with solid size because of the lowered cohesive energy of the under-coordinated surface atoms and the size dependent portion of surface atoms of the entire specimen.[7] $T_C$ elevation by reducing the solid size happened to nanoclusters consisting of III-A and IV-A elements. At the lower end of the size limit, the melting points of Sn, Pb, and Al clusters containing several tens of atoms become higher than the bulk values because the bond strengthening upon the coordination number is heavily reduced.[8] The $T_C$ depression and elevation are also called undercooling or overheating, respectively.

Overheating does occur when the measurement is conducted under applied pressure that compresses the bond length and strengthens the bonds. When the pressure is increased to a critical value of $P_C$, solids transform from the less-densely packed geometry to denser solid phases disregarding the solid size. Conventionally, studies of phase transitions in condensed phases have generally assumed that the stimuli of pressure, temperature, and the composition of the specimen are the variables of dominance in determining the stable states of a substance. The phase transition is well described using the classical theory of statistic thermodynamics in terms of Gibbs free energy without needing consideration of atomistic origin. However, the introduction of the new degree of the freedom of solid size has proven to tune the physical extent of a material, adding another important variable in determining the transition behavior.[9,10] At the nanometer regime, both the thermodynamics and kinetics may come into play in the transformation,[11] which goes beyond the scope of classical theory as the entropy, for instance, is based on statistics of large sample size. Furthermore, all the detectable properties such as the Debye temperature, the specific heat, and the thermal expansion coefficient remain no longer constant but become tunable with the particle size.[12] Despite the importance of the solid-solid transition of materials, the microscopic mechanisms of the solid-solid phase transition of solid in the nanometer regime are thought much more complicated due to the inhomogeneous kinetic effect.[13]

Studies of clusters in the condensed phases have led to numerous instances in which one bond-geometry appears to be favored over others in finite size, as compared to the bulk,[14] because of the altered atomic cohesive energy upon solid size change. Since the pioneer works of Alivisatos and co-workers[10,11,15,16,17,18,19,20,21,22,23,24] there has been a huge database showing consistently that the critical pressure for the solid-solid transition from the less-coordinated structural phase to the denser structures increases with the reduction of crystal size. For the bulk CdSe, the transition pressure is 2.5 GPa but when the size is reduced to 1 to 3 nanometers across, the transition pressure increases to a value of 5 GPa. The size trend for the pressure-induced $\gamma$-$Fe_2O_3$ (maghemite) to $\alpha$-$Fe_2O_3$ (haematite) transition showed that 7 nm nanocrystals transforms at 27 GPa, 5 nm ones at 34 GPa, and 3 nm ones at 37 GPa.[10]

Numerous theoretical approaches have been developed for the possible mechanism of the solid-solid phase transition from various perspectives.[13,20,25,26,27,28,29,30,31] However, finding the factors ruling the size trend of the pressure-induced phase transition in nanocrystals and theoretically reproduction of the observed trends have long been high challenges. The coupling of solid size, temperature, and pressure has led to new phenomena requiring clear understanding. Here we show that an extension of the recently developed bond-order-length-strength (BOLS) correlation mechanism[1] to the pressure domain has led to an analytical expression for the correlation between the critical pressure and the particle size and their effect on the phase stability.

II      Principle

2.1 Atomic cohesive energy and phase stability

It has been clear[7,8] that the factor controlling the thermal stability at a specific atomic site is the atomic cohesive energy that is the sum of bond energy over all the coordinates of the specific atom. If the cohesive energy of a specific atom is higher than the bulk value, overheating occurs to this specific site; otherwise, undercooling takes place. Therefore, it is possible to tune the thermal stability by varying the bond energy or the coordination of the representative bonds or their average over the entire specimen. The $T_C$ of the specimen correlates to the mean atomic cohesive energy that is the sum of bond energy over all the coordinates ($z_i$) of all the $N_j$ atoms of a given specimen of $K_j$ radius:

$$<T_C(K_j)> \propto\, <\sum_{i=1}^{N_j} z_i E_i>$$

(1)

$K_j$ being the dimensionless form of size ins the number of atoms lined along the radius of a spherical specimen. The proportional relation means that a particular process of phase transition costs a certain portion of the cohesive energy. For instance, the liquid-vapor transition costs 100% of the cohesive energy as we need to break all the bonds to evaporate the liquid into individually isolated atom. The process of melting costs only a portion of the cohesive energy because the loosening of the bonds in the melting process. As we focus on the relative $T_C$ change, the exact portion for the specific process will not come into play.

2.2     Local bond average and the pressure-induced $T_C$ elevation

For a given specimen, the bond nature and the total number of bonds do not change before the phase transition taking place. However, the bond length and bond strength will response to the external stimulus such as coordination environment, temperature, and pressure. Therefore, we can focus on the representative bonds or their average to approach the behavior of the entire specimen. Using this approach of local bond average, we are free from considering the concepts such as surface stress, surface energy, and the entropy, as implemented in the classical theories of thermodynamics.

If the applied pressure is increased, all the bonds of the specimen become shorter and stronger because of the volume shrinkage and deformation energy storage. It is reasonable to assume that the pressure-induced energy storage is equally distributed to all the bonds without needing discrimination of the bonds in the surface skin from the ones in the core interior. In fact, in the surface skin, the elastic modulus is relatively higher than the bulk interior because of the high energy density in the surface skin.[32] As the inverse of modulus, the compressibility in the surface skin is lower. The bonds in the surface skin are less compressible compared with the bonds in the bulk though the extent of bond compression at various sites may not be readily detectable at this moment. A recent work[33] using extended x-ray absorption fine structure (EXAFS) analysis on the pressure-induced $LiV_2O_4$ phase transition revealed that the first RDF peak is less sensitive to the applied pressure. Nevertheless, taking the average over all the bonds or over all the atoms of the entire specimen renders no physical indication of the approach.

The pressure-induced energy storage equals to the integral of the volume-pressure (*V-P*) profile shown in Figure 1 from P = 0 to the $P_C$,

$$\Delta E(P_C) = \int_0^{P_C} V dp - V_C P_C = V_0 \int_0^{P_C} \left(1 + \int_0^P \beta dp \right) dp - V_C P_C$$
$$\approx V_0 P_C (1 + \beta P_C /2) - V_C P_C = \sum_{i=1}^{N_j} z_i (\Delta E_i)$$
$$\propto < \Delta E_i > \propto \Delta T_C (P_C)$$

(2)

As it can be seen from the V-P profile in Figure 1, the slope $\beta = dV/(V_0 dp)$ can be taken as a constant for simplicity in numerical calculations. The linear approximation of the $\beta$ leads to the simple relation of $V = V_0 \left(1 + \int_0^P \beta dp \right) \cong V_0 (1 + \beta P)$. $V_C$ and $P_C$ are the critical volume and the critical pressure, respectively. The area of $V_C P_C$ in Figure 1 does not contribute to the cohesive energy. According to eq (2), the mean $T_C$ change of the specimen will increase linearly with the mean bond energy increment if the local bond average is applied. Therefore, the pressure-induced elevation of the $T_C$ for a crystal of $K_j$ radius, $T_C(P_{cb})$, follows the relation,

$$\frac{T_C(P_{Cj}) - T_C(P_{Cb})}{T_C(P_{Cb})} = \frac{<E_i(P_{Cj})> - <E_i(P_{Cb})>}{<E_i(P_{cb})>}$$
$$= \frac{\int_0^{P_{Cj}} \left(1 + \int_0^P \beta dx \right) dp - P_{Ci} V_{Ci}/V_0}{\int_0^{P_{Cb}} \left(1 + \int_0^P \beta dx \right) dp - P_{Cb} V_{Cb}/V_0} - 1 \cong \frac{P_{Cj}(1 + \beta P_{Cj}/2 - V_{Ci}/V_0)}{P_{Cb}(1 + \beta P_{Cb}/2 - V_{Cb}/V_0)} - 1$$

(3)

Eq (3) corresponds to the area ratio between the *V-P* integration from the bulk $P_{Cb}$ to the critical pressure $P_{Cj}$ for the jth particle and the *V-P* integration from P = 0 to the bulk $P_{Cb}$.

2.3    BOLS correlation and the sized induced $T_C$ change

The size dependence of nanostructures in many aspects follows the BOLS correlation mechanism.[1] The BOLS correlation indicates that if one bond breaks, the remaining neighboring ones will be shorter and stronger. The broken bonds induced bond strain and bond strength gain contribute to the atomic cohesive energy of the under-coordinated atoms in the surface skin. Therefore, the size effect on the $T_C$ is dominated by the under-coordinated atoms in the skin shells yet atoms in the core interior remain their bulk nature making no contribution to the $T_C$ change. According to the BOLS correlation and the core-shell configuration of a spherical nanocrystal containing $N_j$ atoms,[1] we have the size dependent $T_C$ of the nanocrystal,

$$\frac{\Delta T_c(K_j)}{T_c(\infty)} = \frac{N_j z_b E_b + \sum_{i \le 3} N_i (z_i E_i - z_b E_b)}{N_j z_b E_b} - 1$$

$$= \sum_{i \le 3} \frac{3C_i}{K_j}(z_{ib} C_i^{-m} - 1) = B/K_j$$

$$\begin{cases} C_i = 2/\{1 + \exp[(12 - z_i)/(8z_i)]\}, & (bond-contraction-coefficient) \\ E_i = C_i^{-m} E_b & (bond-energy-response) \\ N_i/N_j = 3C_i/K_j & (sureace-to-volume-ratio) \end{cases}$$

(4)

where $T_c(\infty)$ is the transition temperature for the bulk $K_j = \infty$. $N_i$ is the number of atoms in the i th atomic shell that is counted from the outermost atomic layer to the center of the solid. The index m = 4 is the bond nature indicator for alloys and compounds. $z_{ib} = z_i/z_b$ is the relative coordination number of an atom in the ith atomic layer. The $z_b$ takes the standard bulk value of 12 without needing to consider the exact crystal structure or orientation for the first order approximation, which leads to only slight deviation in numerical solutions from the true situation. The effective atomic coordination varies with the particle radius in the relation of $z_1 = 4(1-0.75/K_j)$, $z_2 = z_1+2$, and $z_3 = 12$.[1] The B = $\sum_{i \le 3} 3C_i (z_{ib} C_i^{-m} - 1) = -2.56$ can be calculated by taking m = 4. More details about the parameterization please refer to refs [1,7,8]. The sum is over the outermost two or three atomic layers of the nanoparticle as no apparent bond order loss occurs to atoms at i >3, which means that the size-induced $T_C$ suppression arises only from the skin of two or three atomic layers, as confirmed recently by Slezak et al[34] using nuclear inelastic scattering of synchrotron radiation and Gilbert et al[35] using wide-angle x-ray scattering and extended x-ray edge absorption fine structures analyses. Slezak et al found that atoms of the outermost layer of a nanocrystal vibrate with frequencies that are significantly lower and with amplitudes that are much larger than those in the bulk. The vibrations of the second layer are already very close to those of the bulk, being consistent to the assumption made by Jiang[2] and Shi.[3] Gilbert et al[35] found that a structural coherence loss occurs to a 3.4 nm sized ZnS nanoparticle over a distance of 2 nm rather than the diameter of the crystal. The peak positions of the pair distribution function are shifted closer to the reference atom. The shift is more apparent for the nearest and the next, being shortened by 0.008 and 0.02 nm, respectively, indicating the bond contraction dominating in the first and the second atomic shells.

III     Results and discussion
    3.1 The V-P profiles

The typical V-P profiles[36] measured at the ambient temperature for SnO$_2$ bulk and powders of 14, 8, and 3 nm across in Fig. 1 show clearly the size dependence of the critical pressure for phase transition. The $P_C$ is higher for the smaller solid. The V-P profile could be well represented by the relation of $V/V_0 = 1+\beta P$ at $V/V_0 > 0.8$, being substantially the same in value to the description of the third order Birch-Murnaghan equation of state.[37] The slope of the V-P profile, $\beta = dV/(V_0 dp)$, corresponds to bulk compressibility. The $\beta$ values derived from the V-P profile at different regions change slightly, as listed in Table 1. As the $\beta$ is the inverse of bulk modulus, the slight lowered $\beta$ value in the size range of 8 and 14 nm corresponds to turning point of the inverse Hall-Petch relationship for the size dependence of the mechanical strength of nanostructures.[38] By nature, the mechanical strength and the elastic modulus of a specimen are both intrinsically proportional to the sum of bond energy per unit volume though artifacts may be involved in the plastic deformation. Therefore, the lowered $\beta$ in the size region of 8-14 nm is consistent with the strongest size in the inverse Hall-Petch relationship.

For the SnO$_2$ instance, we can estimate the value of $P_{Cb}(1+\beta P_{Cb}/2) = 22.3$ GPa by using eq (3), which is the energy density (being the same in dimension to pressure) stored into the bulk SnO$_2$ prior to the phase transition occurring at 23 GPa. For the 14-nm and 8-nm sized SnO$_2$ samples, the $\Delta T_C(P_{cj})/T_C(P_{cb})$ values are calculated as 0.242 and 0.294, respectively. An extrapolation of the theoretical curve in Figure 2 leads to an estimation of the $\Delta T_C(P_{cj})/T_C(P_{cb})$ value of 0.611 for the 3 nm sized crystal as denoted with the empty square in Figure 2.

3.2  Interdependence of pressure and size

Combining the pressure-induced T$_C$ elevation, eq (3), and the size induced T$_C$ depression, eq (4), we have the immediate relation for the K$_j$ and P$_{Cj}$ interdependence:

$$\frac{\Delta T_C(K_j, P_{Cj})}{T_C(\infty, P_{Cb})} = \frac{\Delta T_C(P_{Cj})}{T_C(P_{Cb})} + \frac{\Delta T_C(K_j)}{T_C(\infty)} = \frac{P_{Cj}(1+\beta P_{Cj}/2 - V_{Ci}/V_0)}{P_{Cb}(1+\beta P_{Cb}/2 - V_{Cb}/V_0)} - 1 + \frac{B}{K_j} = \delta$$

or,

$$K_j = \frac{B}{1+\delta - P_{cj}/A_j} \overset{\delta=0}{=} \frac{BA_j}{A_j - P_{Cj}}; or,$$

$$P_{Cj} = A_j\left(1 - \frac{B}{K_j} + \delta\right) \cong P_{Cb}\left(1 - \frac{B}{K_j} + \delta\right)$$

$$A_j = P_{Cb}(1+\beta P_{Cb}/2 - V_{Cb}/V)/(1+\beta P_{Cj}/2 - V_{Ci}/V_0) \leq 1.03 P_{Cb} \sim P_{Cb}$$

(5)

with A$_j$ value being obtained from the integration of the *V-P* profile in different regions, as listed in Table 1. Note that A$_j$ is slightly size dependence. Within the considered sizes of 3nm or larger, the maximal A$_j$ is 103% of the P$_{Cb}$. For simplicity, one can take A$_j$ ~ P$_{Cb}$ in practice because the 3% difference is within the error of measurement.

The derived expression (5) may serve as a general rule for the size, pressure, and temperature dependence of phase transition, being free from freely adjustable parameters or concepts used in classical theories of thermodynamics. If the T$_C$ values for all the possible sizes are measured different, $\delta \neq 0$. For the phase transitions occurred at the identically ambient temperature, $\delta = 0$.

Figure 2 shows the consistence between predictions (solid curves, eqs (3)) and the measured (scattered symbols) size and pressure dependence of T$_C$ change. The T$_C$ for the solid-solid transition of SnO$_2$ nanocrystals increases under the compressive stress when the solid size is reduced. The T$_C$ for the solid-liquid transition of CdS nanocrystals drops with solid size at the ambient pressure.[39] The exceedingly good agreement between predictions and observations in the two situations evidences the validity of the considerations from the perspective of atomic cohesive energy. The critical pressure for the 3 nm $SnO_2$ transition is predicted to be 43 GPa that is beyond the range of the referred measurement. It is clear now that the pressure-induced $T_C$ elevation compensates for the size-induced $T_C$ depression, which determines the observed size trend of the pressure induced phase transition. If the solid size induces a $T_C$ elevation,[40] the critical pressure will drop to compensate for the size-induced overheating. Therefore, the critical pressure for phase transition at a given temperature may increase or decrease subjecting to the size effect.

With the relation given in eq (5), we are able to reproduce the measured $P_c$-$K_j$ data for CdSe[11,15,16] transforming from the 4-nearest coordinated Wurtizite to the 6-nearest coordinated cubic rock-salt structure, and the transition of $Fe_2O_3$ [10] from γ to α phase, as shown in Figure 3. In calculations, the input for the CdSe nanocrystal are only the bulky β value of 0.0169 $GPa^{-1}$ and the bulk critical pressure of 2.5 GPa.[41] For the $Fe_2O_3$ specimens, we used the $A_j = P_{Cb}$ approximation without needing the input of β value. The exceedingly good agreement between the predictions and the measured trends in both cases demonstrates the effectiveness of eq (5) and the validity of the physical considerations.

IV  Conclusion

In summary, we have developed an analytical form for the size dependence of the pressure-induced phase transition in nanocrystals from the perspective of bonding energetics and its response to external stimulus such as coordination environment, pressure, and temperature. We found that the competition between the size-induced undercooling and the pressure-induced overheating dominates the change yet no concepts in classical theories of thermodynamics such as surface stress or surface energy are necessary. Reproduction of the measured data evidences that the derived expression and the approaches of local bond average and the BOLS correlation could represent the true situations of observations that may go beyond the scope of classical and quantum approaches. The developed form could serve as a rule for predicting the size dependence of the pressure- and temperature-induced phase transition in nanostructures.

The project is supported by NSF of China (Nos. 10525211 and 10772157) and MOE (RG14/06), Singapore.

Table & Figure captions

Figure 1 The *V-P* profile for $SnO_2$ nanocrystals phase transformation from rutile to cubic structures.[36] The square, diamond, circle and up-triangle symbols represent bulk, 14-nm, 8-nm, and 3-nm $SnO_2$ samples, respectively. Corresponding critical transition pressures are denoted as $p_{Cb}$, $p_{C1}$, and $p_{C2}$ while the 3 nm crystal case is beyond the measured pressure range. The slopes correspond directly to the compressibility or the inverse of bulk modulus. The integration of the *V-P* profile represents energy stored in the crystal to raise the transition temperature.

Figure 2 Comparison of the predicted (solid curves) with the measured (scattered symbols) pressure-induced $T_C$ elevation for solid-solid transition of $SnO_2$ nanocrystals derived from Figure 1 and the size-induced $T_C$ depression for solid-liquid transition of CdS nanocrystals, showing compensation of the relative changes.[39] The empty square is an extrapolation of the 3nm $SnO_2$ solid. The transition pressure for the 3 nm $SnO_2$ is predicted to be 43 GPa.

Figure 3 Theoretical reproduction (curves) of the measured (scattered data) $P_C$-R relations for (a) CdSe [11,15,16] and (b) $Fe_2O_3$ [Ref 10] nanostructures. R is the radius of the nanocrystal.

Table 1 Information derived from Figure 1 for SnO$_2$ nanocrystals. $\beta_j$ is the slope of the V-P profile, $\Delta S_j$ the integration of the V-P curve. The meaning of A$_j$ and $\Delta T_C(P_{cj}, K_j)/T_C(P_{cb}, \infty)$ are given in the context.

|  | bulk | 14 nm | 8 nm | 3 nm |
|---|---|---|---|---|
| $\beta_j$ (10$^{-3}$ GPa$^{-1}$) | -3.53 | -2.35 | -2.35 | -3.3 |
| $\Delta S_j$ (GPa) | 22.26 | 5.39 | 6.54 | 10.38 |
| P$_{cj}$(GPa) | 23.0 | 29.0 | 31.0 | 43.0 |
| A$_j$(GPa) | 23.0 | 23.46 | 23.47 | 23.49 |
| $\Delta T_C(P_{cj}, K_j)/T_C(P_{cb}, \infty)$ | 0 | 0.242 | 0.294 | 0.611 |

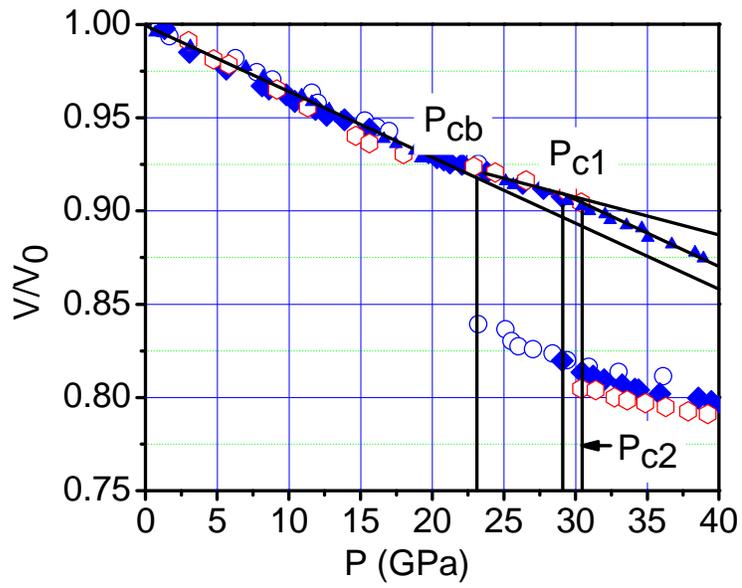

Figure 1

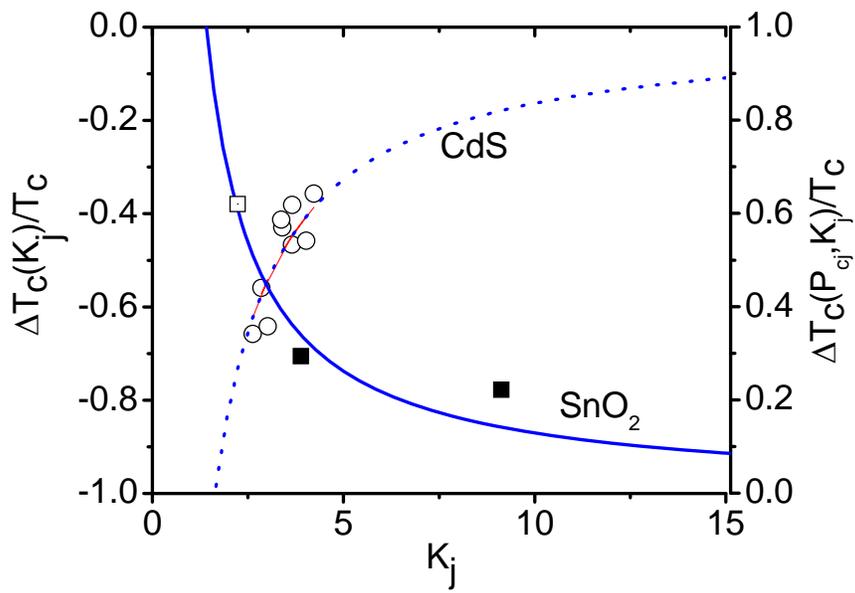

Figure 2

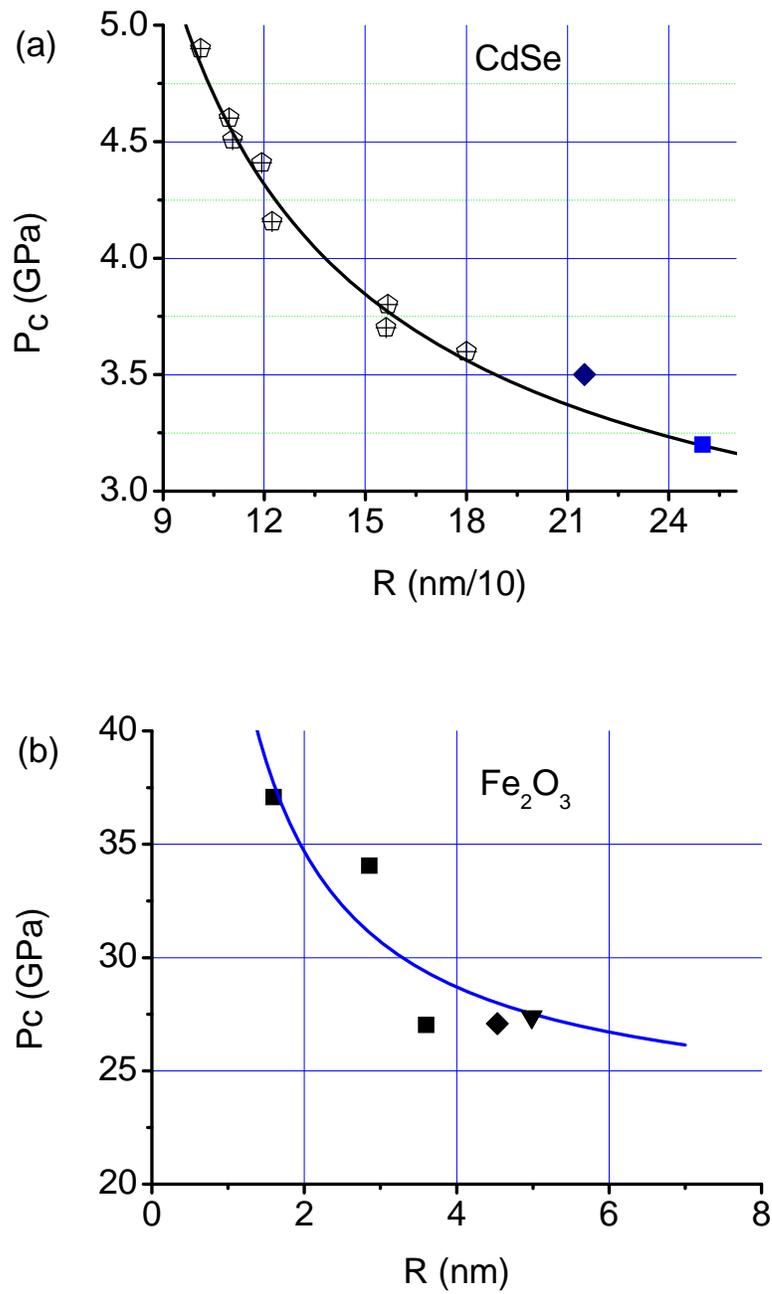

Figure 3

---

[1] Sun, C. Q.; *Prog Solid State Chem* **2007**, *35*, 1-159.
[2] Zhang, Z.; Li, J. C.; Jiang, Q. *Journal of Physics D-Applied Physics* **2000**, *33*, 2653.
[3] Shi, F. G. *J. Mater. Res.* **1994**, *9*, 1307.